\documentclass[preprintnumbers,showpacs]{revtex4}
\usepackage{graphicx}
\usepackage{color}

\def\bc{\begin{center}}
\def\ec{\end{center}}

\def\beq{\begin{equation}}
\def\eeq{\end{equation}}

\begin{document}

\title{Towards superfluidity of  dipolar excitons in a TMDC double layer}
\author{Oleg L. Berman$^{1,2}$ and Roman Ya. Kezerashvili$^{1,2}$}
\affiliation{\mbox{$^{1}$Physics Department, New York City College
of Technology, The City University of New York,} \\
Brooklyn, NY 11201, USA \\
\mbox{$^{2}$The Graduate School and University Center, The
City University of New York,} \\
New York, NY 10016, USA}
\date{\today}

\begin{abstract}

We study formation and superfluidity of dipolar excitons in double
layer heterostructures formed by two transition metal dichalcogenide
(TMDC) atomically thin layers. Considering screening effects for an
electron-hole interaction via the harmonic oscillator approximation
for the Keldysh potential, the analytical expressions for the
exciton energy spectrum and the mean field critical temperature
$T_{c}$ for the superfluidity are obtained. It is shown that binding
energies of A excitons are larger than for B excitons. The mean
field critical temperature for a two-component dilute exciton system
in a TMDC double layer is analyzed and shown that latter is an
increasing function of the factor $Q$, determined by the effective
masses of A and B excitons and their reduced mass. Comparison of the
calculations for $T_{c}$  performed  by employing the Coulomb and
Keldysh interactions demonstrates  the importance of screening
effects in TMDC.

\end{abstract}


\maketitle

\section{Introduction}

\label{intro}

When a sufficient amount  of  bosons at low temperatures
spontaneously occupy the single lowest energy quantum state,
Bose--Einstein condensation (BEC) happens. The system of interacting
bosons can experience the superfluidity, caused by BEC, analogously
to the superfluid helium~\cite{Pitaevskii}. A BEC of weakly
interacting particles was achieved experimentally in dilute gases of
alkali atoms. This atomic BEC can be created at the nanokelvin
temperatures, which are technically challenging to achieve.  The progress
in the experimental and theoretical research  of the BEC of dilute
supercold alkali gases is reviewed in Ref.~\cite{Daflovo}.

The BEC for two-dimensional (2D) bosons with higher mass occurs at
the temperatures greater than for bosons with lower mass, because
the de Broglie wavelength for 2D system is inversely proportional to
the square root of the mass of a particle. Therefore,  BEC exists at
much higher temperatures in  a Bose gas of small mass particles,
than  in a system of relatively heavy alkali atoms. The small mass
boson quasiparticles can be formed due to absorption of a photon by
a semiconductor.  Absorption of a photon leads to creation of an
electron in a conduction band and a positive charge ``hole'' in a
valence band. This electron-hole pair can form a bound state called
 an ``exciton''.  The BEC and superfluid  of such excitons are  expected to exist
 at experimentally  observed exciton densities at temperatures much higher than for
the BEC of alkali atoms~\cite{Moskalenko_Snoke}. The BEC and
superfluidity of dipolar (indirect) excitons, formed by electrons
and holes, spatially separated in two parallel two-dimensional
semiconductor quantum wells,  were proposed~\cite{Lozovik}. The
experimental observation of superfluidity of dipolar  excitons in GaAs quantum
wells was claimed recently~\cite{Dubin}. The recent progress in
experimental and theoretical studies of BEC of dipolar excitons in
semiconductor quantum wells was
reviewed~\cite{Moskalenko_Snoke,Butov_rev,Snoke_rev,Combescot}.

Due to relatively large exciton binding energies in novel 2D
semiconductors, such as  transition metal dichalcogenides (TMDC),
the BEC and superfluidity of dipolar excitons in double layers of
TMDC can occur.   Monolayers of TMDC such as
$\mathrm{Mo S_{2}}$, $\mathrm{Mo Se_{2}}$, $\mathrm{Mo Te_{2}}$, $\mathrm{W S_{2}}$, $%
\mathrm{W Se_{2}}$, and $\mathrm{W Te_{2}}$ are 2D semiconductors,
belonging to  a class of monolayer direct bandgap materials, attract an interest due various  applications in electronics and
opto-electronics~\cite{Kormanyos}.  Since contrary to gapless
graphene, TMDC monolayers have the direct gap in a single-particle
spectrum exhibiting the semiconducting band
structure~\cite{Mak2010,Mak2012,Novoselov,Zhao}, excitons in TMDCs
can be created by the laser pumping.   The ground and excited states
of direct excitons in mono- and few-layer TMDCs on a
$\mathrm{SiO_{2}}$ substrate were experimentally and theoretically
studied~\cite{Chernikov}. Two distinct types of excitons in TMDC
layers, labeled A and B, are formed due to significant spin-orbit
splitting in the valence band~\cite{Reichman}. The excitons of type
A are created by spin-up electrons from conduction and spin-down
holes from valence bands. The excitons of type B are created by
spin-down electrons from conduction and spin-up holes from valence
bands. While the spin-orbit splitting in the valence band is much
larger than in the conduction band, in the valence band  the energy
for spin-down electrons is larger than for spin-up electrons. The
spin-orbit spitting results in the experimentally observed energy
difference between the A and B excitons~\cite{Kormanyos}. Therefore,
A and B excitons form a two-component Bose gas in TMDCs.

 High-temperature superfluidity can be observed for
dipolar excitons in a heterostructure of two TMDC monolayers,
separated by a hexagonal boron nitride ($h$-BN) insulating
barrier~\cite{Fogler}. The
dipolar excitons were observed in heterostructures formed by monolayers of $\mathrm{%
Mo S_{2}}$ and $\mathrm{Mo Se_{2}}$ on a
$\mathrm{Si}-\mathrm{SiO_{2}}$ substrate~\cite{Ceballos} and by monolayers of $\mathrm{%
Mo S_{2}}$ on a substrate constrained by hexagonal boron nitride
 layers~\cite{Calman}. The theoretical study of the
phase diagram of 2D dipolar exciton condensates in a TMDC double
layer was performed~\cite{Macref}. The  high-temperature
superfluidity of the two component Bose gas of A and B dipolar
excitons in a transition metal dichalcogenide double layer was
predicted in Ref.~\cite{BK}. $T_{c}$ for a two-component exciton
system in a TMDC double layer is about one order of magnitude higher
than $T_{c}$ for any one-component exciton system, because for a
two-component system $T_{c}$ depends on the effective reduced mass
of A and B excitons, which is always smaller than the individual
effective mass of A or B exciton~\cite{BK}.

In this paper, we study the superfluidity of two-component dilute
Bose gas of dipolar A and B excitons in different TMDC double
layers. We search the candidates for higher temperature of
superfluidity by comparing the results of the calculations for
various TMDC double layers, formed by two monolayers of the same
TMDC material and two different TMDC monolayers, when the transition
metal atom is replaced by the other transition metal atom (e.g. for
a MoS$_{2}$/WS$_{2}$ heterostructure) or when the chalcogenide atoms
are replaced by the other chalcogenide atoms (e.g. for a
MoS$_{2}$/MoSe$_{2}$ heterostructure). While an  electron and a hole
interact via the Coulomb potential, in general, affected by
screening effects the electron-hole interaction in TMDC materials is
described by the Keldysh potential ~\cite{Reichman}. In the
framework of the harmonic oscillator approximation for the
electron-hole interaction in TMDC double layer heterostructure we
obtain the analytical expressions for the exciton energy spectrum
and the mean field critical temperature of superfluidity. The
calculations are performed for both the Keldysh and Coulomb
potentials, describing the interactions between the charge carriers,
which allows to study the influence of the screening effects on the
properties of a weakly interacting Bose gas of dipolar excitons in a
TMDC double layer.

The paper is organized in the following way.  In Sec.~\ref{tbp}, the
two-body problem for an electron and a hole, spatially separated in
two parallel TMDC monolayers, is formulated, and the
 energy spectrum,  wave functions, effective masses and
binding energies  for a single dipolar exciton   in a TMDC  double
layer are obtained. The exciton-exciton interaction is analyzed in
Sec.~\ref{exexint}.  In Sec.~\ref{sup}, the mean field critical
temperature of superfluidity for the two-component dilute system of
dipolar excitons in a TMDC double layer is obtained and analyzed.
The conclusions follow in Sec.~\ref{conc}.


\section{Two-body problem for an electron and a hole, spatially separated in two parallel TMDC monolayers}
\label{tbp}

We consider electrons to be
 confined in a 2D TMDC  monolayer, while an
equal number of positive holes are placed   in a parallel TMDC
monolayer at a distance $D$ away. The system of the charge carriers
in two parallel TMDC  layers is treated as a 2D system without
interlayer hopping. The electron and hole via electromagnetic
interaction $V(r_{eh})$, where $r_{eh}$  is a distance between the
electron and hole, could form a bound state, i.e., a dipolar
exciton. The electron-hole recombination due to the tunneling of
electrons and holes between different TMDC monolayers is suppressed
by the dielectric barrier with the dielectric constant $\varepsilon
_{d}$ that separates two TMDC monolayers~\cite{BK}. Therefore, the
dipolar excitons, formed by electrons and holes, located in two
different parallel TMDC monolayers, have a longer lifetime than the
direct excitons. After projection  the electron position vector onto
the TMDC plane with holes and replacing  the relative coordinate
vector ${\bf r}_{eh}$ by its projection $\mathbf{r}$ on this plane
and taking into account that $r_{eh} = \sqrt{r^{2}+ D^{2}}$, the
potential $V(r_{eh})$ can be expressed as $V(r)=
V(\sqrt{r^{2}+D^{2}}),$ where $r$ is the relative distance between
the hole and the projection of the electron position onto the TMDC
monolayer with holes. Thus, a dipolar exciton can be described by a
two-body 2D Schr\"{o}dinger equation with potential
$V(\sqrt{r^{2}+D^{2}})$, employing in-plane coordinates
$\mathbf{r}_{1}$ and $\mathbf{r}_{2}$ for the hole and the
projection of the electron, respectively, so that $\mathbf{r}=$
$\mathbf{r}_{1}-\mathbf{r}_{2}$.

The Hamiltonian of an electron and a hole, spatially separated in
two parallel TMDC monolayers with the interlayer distance $D$ has
the following form
\begin{equation}
\hat{H}_{ex}=-\frac{\hbar ^{2}}{2m_{e}}\Delta
_{\mathbf{r}_{1}}-\frac{\hbar ^{2}}{2m_{h}}\Delta _{\mathbf{r}_{2}}
+ V (r)  , \label{rk2}
\end{equation}%
where $\Delta _{\mathbf{r}_{1}}$ and $\Delta _{\mathbf{r}_{2}}$ are
the Laplacian operators with respect to the components of the vectors $\mathbf{r}%
_{1}$ and $\mathbf{r}_{2}$, respectively, and $m_{e}$ and $m_{h}$
are the effective masses of the electron and hole, respectively. The
problem of the in-plane motion of interacting electron and hole
forming the exciton in a TMDC double layer can be reduced to that of
one particle with reduced mass $\mu$ in a $V(r)$ potential and
motion of the center-of-mass of the exciton.

After introducing the coordinates of the center-of-mass $\mathbf{R}$
of an exciton  and the coordinate of the relative motion
$\mathbf{r}$ of an electron and hole as
\begin{equation}
\label{cent} \mathbf{R} = \frac{m_{e}\mathbf{r}_{1} +
m_{h}\mathbf{r}_{2}}{m_{e}+ m_{h}} \ , \hspace{2cm} \mathbf{r} =
\mathbf{r}_{1} - \mathbf{r}_{2}  ,
\end{equation}%
we represent the Hamiltonian $\hat{H}_{ex}$ in the form:
$\hat{H}_{ex}= \hat{H}_{R} + \hat{H}_{r}$ where  the Hamiltonians of
the motion of the center-of-mass $\hat{H}_{R}$ and relative motion
of electron and a hole $\hat{H}_{r}$ are given by
\begin{equation}
\label{sep1} \hat{H}_{R}=-\frac{\hbar^{2}}{2M}\Delta _{\mathbf{R}} \
, \hspace{2cm}  \hat{H}_{r}=-\frac{\hbar^{2}}{2\mu}\Delta
_{\mathbf{r}} + V(r) .
\end{equation}%
In Eq.~(\ref{sep1}) $M$ and $\mu$ are the exciton effective mass and
reduced mass, correspondingly, defined as
\begin{equation}
\label{Mm} M = m_{e} + m_{h}  \ , \hspace{2cm} \mu =
\frac{m_{e}m_{h}}{m_{e}+m_{h}}    .
\end{equation}%
The solution of the Schr\"{o}dinger equation for the center-of-mass
of an exciton $\hat{H}_{R}\psi (\mathbf{R}) = \mathcal{E}\psi
(\mathbf{R})$ is the plane wave $\psi (\mathbf{R}) =
e^{i\mathbf{P}\mathbf{R}/\hbar}$ with the quadratic energy spectrum
$\mathcal{E} = P^{2}/(2M)$, where $\mathbf{P}$ is the momentum of
the center-of-mass of an exciton.



\subsection{The harmonic oscillator approximation of Keldysh
potential for a TMDC double layer}

\label{Keld}

In TMDC layers due to the screening
effects~\cite{Reichman,Prada,Saxena,VargaPRB2016,Kezerashvili2016} the electron-hole attraction has to be described by
the Keldysh potential \cite{Keldysh,Rubio}.
We assume the following form of the Keldysh potential~\cite{Rodin}:
\begin{eqnarray}
\label{Keldysh} V(r_{eh}) = -\frac{\pi k
e^{2}}{\left(\varepsilon_{1} + \varepsilon_{2}\right)\rho_{0}}
\left[H_{0}\left(\frac{r_{eh}}{\rho_{0}}\right)  -
Y_{0}\left(\frac{r_{eh}}{\rho_{0}}\right)  \right]   ,
\end{eqnarray}
where
$k=9\times 10^{9}\ N\times m^{2}/C^{2}$, $H_{0}(x)$ and $Y_{0}(x)$
are Struve and Bessel functions of the second kind of order $\nu=0$,
correspondingly, $\varepsilon_{1}$ and $\varepsilon_{2}$ are the
dielectric constants of the dielectrics, surrounding the TMDC layer,
$\rho_{0}$ is the screening length, defined by $\rho _{0}=2\pi
\zeta/\left[\left(\varepsilon_{1} +
\varepsilon_{2}\right)/2\right]$, where $\zeta$ is the 2D
polarizability. If the electron-hole separation is large, implying
$r_{eh} \gg \rho _{0\text{ }}$, the potential, given by
Eq.~(\ref{Keldysh}),  has the three-dimensional Coulomb tail, while
for small electron-hole distances at $r_{eh} \ll \rho _{0\text{ }}$
it turns to  a logarithmic Coulomb potential for two 2D point
charges. Throughout of this paper we assume that TMDC monolayers are
embedded in the same material with dielectric constant
$\varepsilon_{d}$ and, therefore, we set $\varepsilon_{1} =
\varepsilon_{2} = \varepsilon_{d}$.

Substituting $r_{eh} = \sqrt{r^{2}+ D^{2}}$ into
Eq.~(\ref{Keldysh}), assuming $r\ll D$, and expanding
Eq.~(\ref{Keldysh}) in Taylor series we obtain in the first order
with respect to $\left(r/D\right)^{2}$:
\begin{eqnarray}
\label{expand} V(r) = -V_{0} + \gamma r^{2} ,
\end{eqnarray}
where
\begin{eqnarray}
\label{V0} V_{0} =  \frac{\pi k e^{2}}{2 \varepsilon_{d}\rho_{0}}
\left[H_{0}\left(\frac{D}{\rho_{0}}\right) -
Y_{0}\left(\frac{D}{\rho_{0}}\right)  \right]  ,
\end{eqnarray}
\begin{eqnarray}
\label{gamma} \gamma = - \frac{\pi k e^{2}}{4 \varepsilon_{d}\rho_{0}^{2}D}
\left[H_{-1}\left(\frac{D}{\rho_{0}}\right)  -
Y_{-1}\left(\frac{D}{\rho_{0}}\right)  \right].
\end{eqnarray}

If we use the Coulomb potential, the potential energy of the
electron-hole attraction $V(r)$ is
\begin{eqnarray}  \label{V}
V(r)=-\frac{ke^{2}}{\varepsilon _{d}\sqrt{r^{2}+D^{2}}} .
\end{eqnarray}
Assuming $r\ll D$, we approximate $V(r)$ by the first two terms of
the Taylor series and obtain
\begin{eqnarray}  \label{Vap}
V(r)=-V_{0}+\gamma r^{2} , \ \  \text{ where }
V_{0}=\frac{ke^{2}}{\varepsilon _{d}D}, \ \ \ \ \  \ \ \ \gamma =\frac{ke^{2}}{%
2\varepsilon _{d}D^{3}} .
\end{eqnarray}

\begin{figure}[t]
\begin{center}
\includegraphics[width=8.5cm]{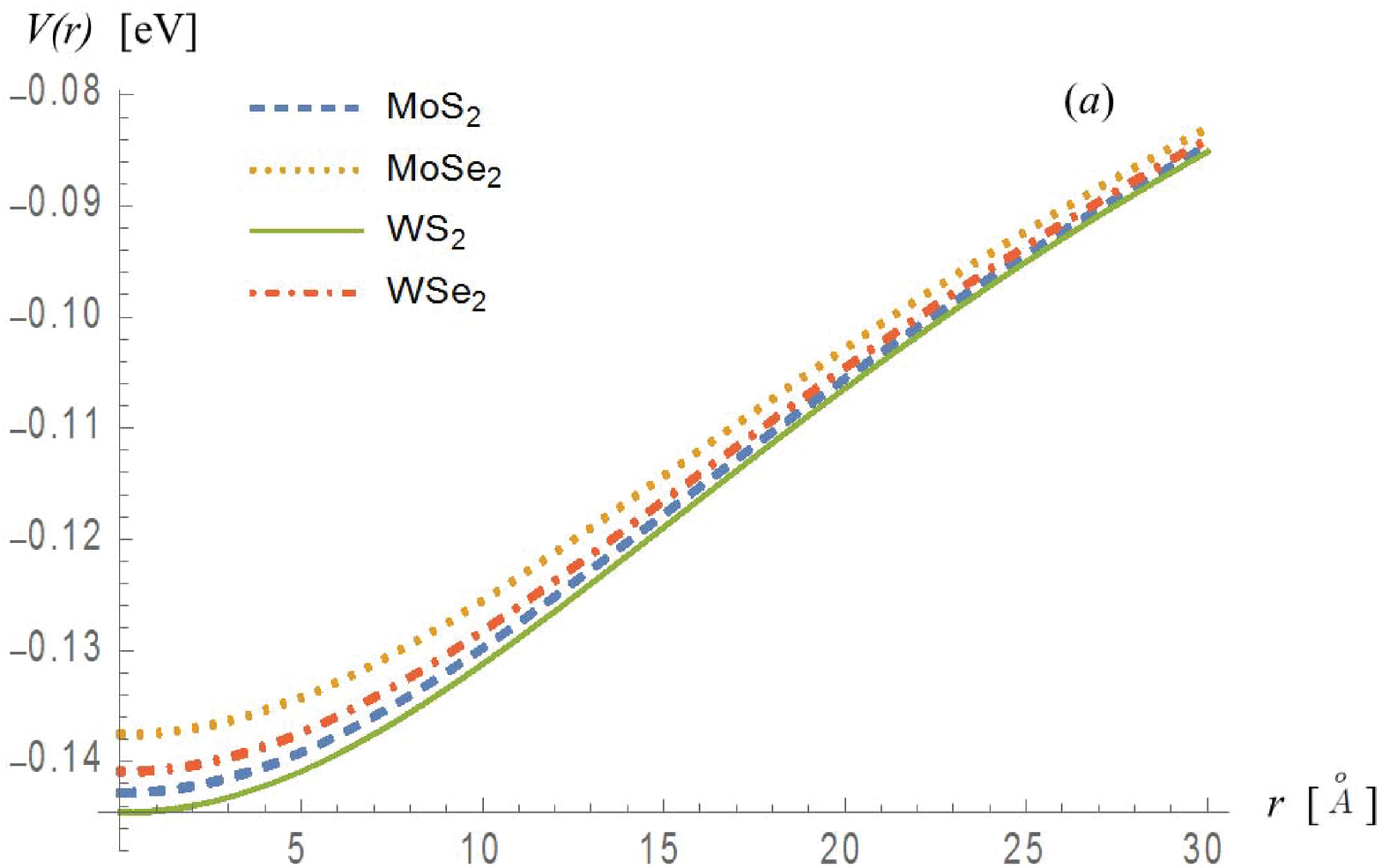} \vspace{-0.1cm} %
\includegraphics[width=8.5cm]{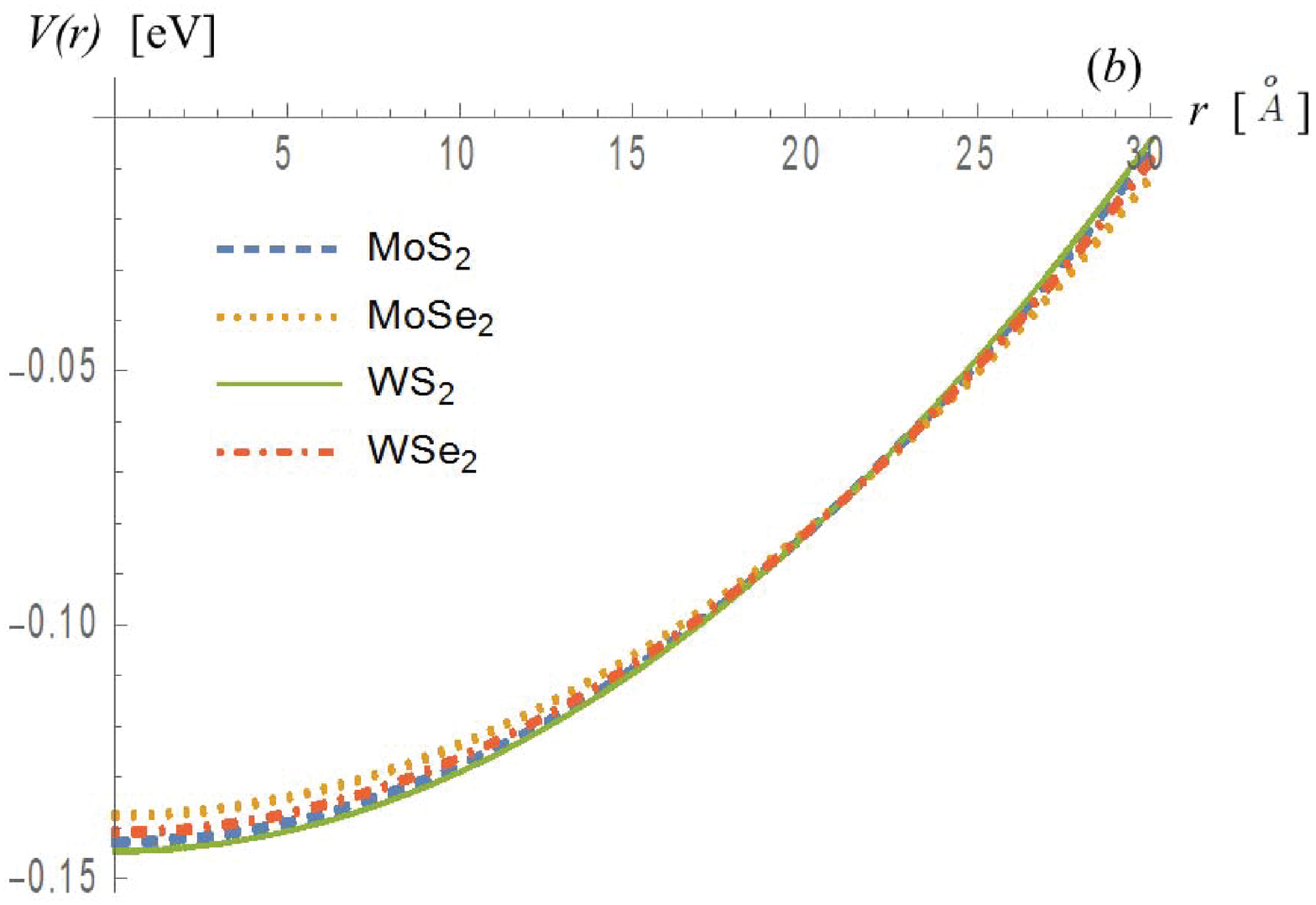} \vspace{-0.1cm} %
\includegraphics[width=8.5cm]{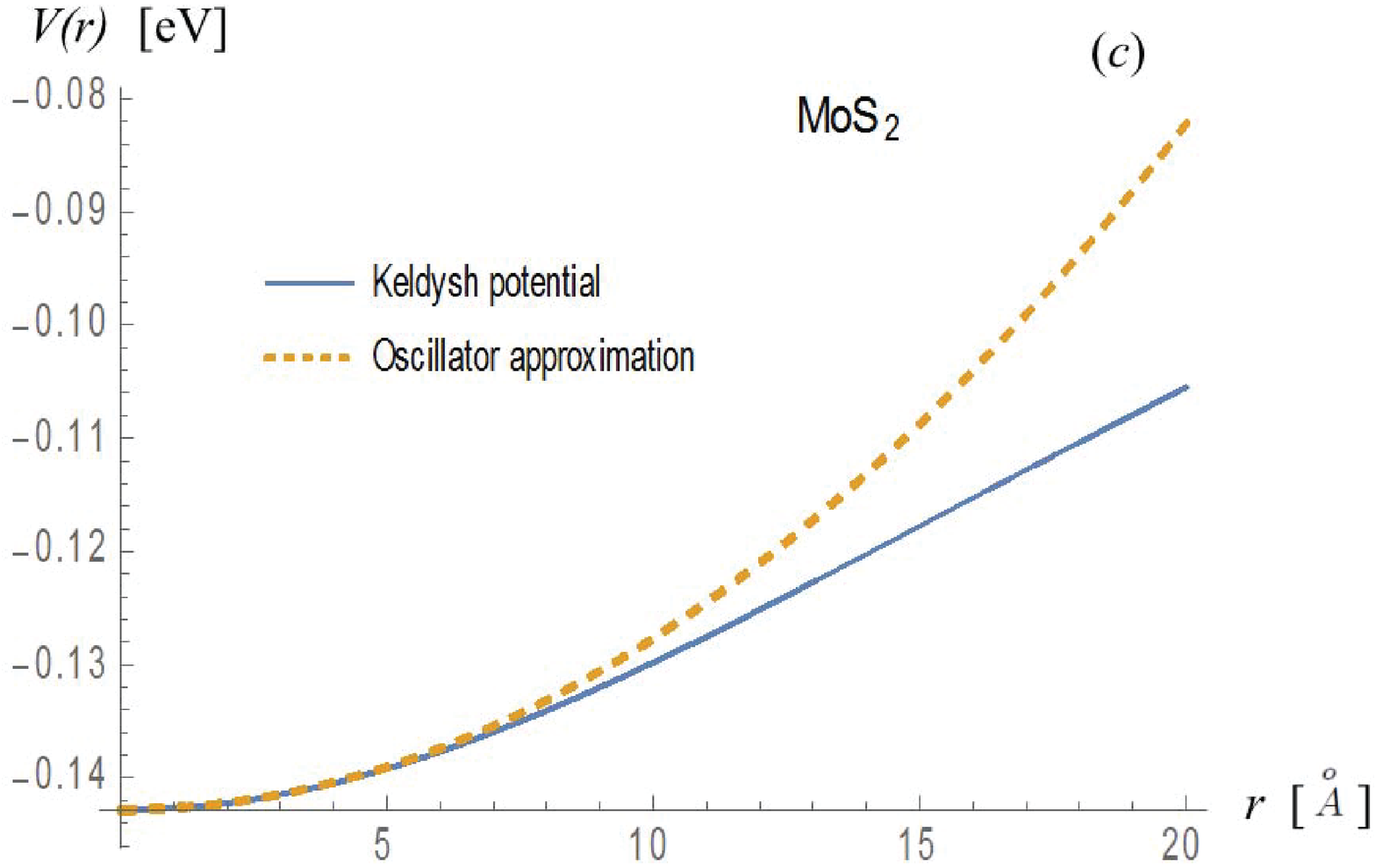} \vspace{-0.1cm}
\end{center}
\caption{(Color online)   (a) The Keldysh  potential for
electron-hole attraction in different TMDC double layers, given by
Eq.~(\ref{Keldysh}). (b) The harmonic oscillator approximation for the Keldysh potential for electron-hole
attraction in different TMDC double layers, given by
Eq.~(\ref{expand}).
(c) Comparison of the Keldysh potential and the Keldysh potential,
approximated by the harmonic oscillator potential in a $\mathrm{Mo
S_{2}}$ double layer. The calculations were performed for the number
$N_{L} = 10$ of $h$-BN monolayers, located between two TMDC
monolayers. The polarizabilities for TMDC materials are taken from
Ref. \protect\cite{hmass}.}
 \label{Fig1}
\end{figure}

 The Keldysh  potential and the harmonic oscillator approximation for the Keldysh potential for electron-hole
attraction in  a TMDC double layer, describing the electron-hole
interaction in  a TMDC double layer, given by Eqs.~(\ref{Keldysh})
and (\ref{expand}), respectively, are shown in Fig.~\ref{Fig1}.
According to Figs.~\ref{Fig1}a and~\ref{Fig1}b, the difference
between the potentials for different TMDC materials decreases as $r$
increases. As seen from Fig.~\ref{Fig1}c, the harmonic oscillator
approximation of the Keldysh potential is very much close to the
exact Keldysh potential for small $r$,
while it becomes larger than the exact Keldysh potential when $r$
increases. Let us mention that indirect excitons were observed in two different
$\mathrm{Mo S_{2}}$  layers separated by $h$-BN
 monolayers  and surrounded by
 $h$-BN  cladding layers, since $h$-BN monolayers are characterized by relatively small density of the defects of their crystal structure~\cite{Calman}.
 Therefore, in our calculations we consider the TMDC monolayers to be separated
by $h$-BN insulating layers. Besides we assume $h$-BN insulating
layers to be located on the top and on the bottom of the TMDC double
layer. In this case, $\varepsilon_{d} = 4.89$ is the effective
dielectric constant, defined as $\varepsilon_{d} =
\sqrt{\varepsilon^{\bot}}\sqrt{\varepsilon^{\parallel}}$~\cite{Fogler},
where $\varepsilon^{\bot}= 6.71$ and $\varepsilon^{\parallel} =
3.56$  are the components of the dielectric tensor for
$h$-BN~\cite{CaihBN}. Throughout of this paper we consider the
separation between two layers of TMDC materials in steps of $D_{hBN}
= 0.333$ nm, corresponding to the thickness of one $h$-BN
monolayer~\cite{Fogler}. Therefore, the interlayer separation $D$ is
presented as $D = N_{L}D_{hBN}$, where $N_{L}$ is the number of
$h$-BN monolayers, placed between two TMDC monolayers. In Fig.
\ref{Fig1} the number $N_{L} = 10$ of $h$-BN monolayers, located
between two TMDC monolayers corresponds to $D = 3.33 \ \mathrm{nm}$.

\subsection{The dipolar exciton energy spectrum and wave functions}

\label{rel}

The solution of the Schr\"{o}dinger equation for the relative motion
of an electron and a hole  $\hat{H}_{r}\Psi (\mathbf{r}) = E\Psi
(\mathbf{r})$ with  the potential (\ref{expand}) is reduced to the
problem of a 2D harmonic oscillator with the exciton reduced mass
$\mu$ defined by Eq.~(\ref{Mm}).
 The eigenfunctions and eigenenergy for a
single particle in the parabolic well were first determined by Fock
in 1928~\cite{Fock}, later in Ref. [\onlinecite{Darwin}], and was
studied in detail in Ref. \cite{Dingle}. The single-particle
eigenfunction for the two-dimensional oscillator  was widely used
for the description of  a quantum dot \cite{Maksym}.

Following Refs.~[\onlinecite{Maksym,Iyengar}] one obtains the radial
Schr\"{o}dinger equation and the solution for the eigenfunctions for
the relative motion of an electron and a hole  in a TMDC double
layer in terms of associated Laguerre polynomials can be written as
\begin{equation}
\Psi_{N L} (\mathbf{r})=\frac{N!}{a^{|L|+1}\sqrt{\widetilde{n}!%
\widetilde{n}^{\prime }!}}2^{-|L|/2}\mathrm{sgn}(L)^{L}r^{|L|}e^{-r^{2}/(4a^{2})}%
\times L_{N}^{|L|}(r^{2}/(2a^{2}))\frac{e^{-iL\phi }}{(2\pi )^{1/2}}
, \label{rk14}
\end{equation}%
where $N=\mathrm{min}(\widetilde{n},\widetilde{n}^{\prime })$, $L=\widetilde{%
n}-\widetilde{n}^{\prime }$, $\widetilde{n},$ $\widetilde{n}^{\prime
}=0,1,2,3,\ldots $ are the quantum numbers, $\phi $ is the polar
angle angle, and $a=\left[ \hbar /\left(2\sqrt{2\mu\gamma
}\right)\right]^{1/2}$.

The corresponding energy spectrum is given by
\begin{equation}
E_{N L} \equiv E_{e(h)} = - V_{0} + (2N+1+|L|)\hbar \left( \frac{2\gamma}{\mu}%
\right) ^{1/2} .  \label{rk15}
\end{equation}
At the lowest quantum state $N = L = 0$ as it follows from
Eq.~(\ref{rk15})  the ground state energy  for the exciton is given
by
\begin{equation}
E_{00}=  - V_{0} + \hbar \left( \frac{2\gamma}{\mu}%
\right) ^{1/2} .  \label{rk16}
\end{equation}

The important characteristics of the exciton is the square of the
in-plane gyration radius $r_{X}^2$. It allows to estimate the
condition when the excitonic gas is dilute enough. One can obtain
the square of the in-plane gyration radius $r_{X}$ of a dipolar
exciton~\cite{Fogler}, which is the average squared projection of an
electron-hole separation onto the plane of a TMDC monolayer as
\begin{equation}
r_{X}^{2} \equiv \left\langle r^{2} \right\rangle = \int
\Psi_{00}^{*} (\mathbf{r}) r^{2} \Psi_{00} (\mathbf{r}) d^{2} r =
\frac{2\pi}{2\pi a^{2}}\int_{0}^{+\infty} r^{2} e^{- \frac{r^{2}}{2
a^{2}}} r d r   = 2 a^{2} .
 \label{rx2}
\end{equation}

Let us mention that the Taylor expansion of the electron-hole
attraction potential in the first order with respect to
$r^{2}/D^{2}$, presented by Eq.~(\ref{expand}) is valid if
$r_{X}^{2} = 2 a^{2} \ll D^{2}$. Thus, one obtains that
$\hbar/\sqrt{2 \mu\gamma} \ll D^{2}$. The latter inequality holds
for  $D \gg D_{0}$. The value of $D_{0}$ depends on the effective
masses of the electron and hole.  For the Keldysh potential $D_{0}$
can be obtained from solution of the following transcendental
equation:
\begin{equation}
D_{0}^{3} = - \frac{2\hbar^{2} \varepsilon_{d}\rho_{0}^{2}}{\pi k
e^{2}\mu\left[H_{-1}\left(\frac{D_{0}}{\rho_{0}}\right)  -
Y_{-1}\left(\frac{D_{0}}{\rho_{0}}\right)  \right] }  .
\label{KeldD0}
\end{equation}
The following inequality should hold for the Keldysh potential:
\begin{equation}
- \frac{2\hbar^{2}\varepsilon_{d}\rho_{0}^{2}}{\pi k
e^{2}\mu\left[H_{-1}\left(\frac{D}{\rho_{0}}\right) -
Y_{-1}\left(\frac{D}{\rho_{0}}\right)  \right] } \ll D^{3}  .
\label{KeldD02}
\end{equation}
For the Coulomb potential, we have $D_{0} =
\hbar^{2}\varepsilon_{d}/ \left(ke^{2}\mu\right)$. The comparison of
the latter expression for $D_{0}$ with Eq.~(\ref{KeldD0}) shows that
in the case of the Keldysh potential $D_{0}$ depends on the
screening length which is  contingent on the 2D polarizability of
TMDC material.

To estimate the condition of validity of the Taylor expansion we
find $D_{0}$ by solving Eq.~(\ref{KeldD0}). The masses of dipolar
excitons and $D_{0}$ for different TMDC materials are represented in
Table~\ref{tab6}. The smallest $D_{0}$ corresponds to a MoSe$_{2}$
double layer, while largest $D_{0}$ corresponds to a WS$_{2}$ double
layer. The values of $D_{0}$ are smaller for $A$ dipolar excitons
than for B dipolar excitons for the same TMDC double layers.
Consideration of the double layer heterostructure that consists from
two different TMDC layers, when in one of the layers the transition
metal atom is replaced by the other transition metal atom (e.g. for
a MoS$_{2}$/WS$_{2}$ heterostructure) or the chalcogenide atoms are
replaced by the other chalcogenide atoms (e.g. for a
MoS$_{2}$/MoSe$_{2}$ heterostructure) changes the value of $D_{0}$
insignificantly.

\begin{table}[t]
\caption{Masses of excitons and $D_{0}$ for different TMDC
materials. $m_{0}$ is an electron mass.}
\begin{center}

\begin{tabular}{cccccc}
\hline\hline
Exciton &  & MoS$_{2}$ & MoSe$_{2}$ & WS$_{2}$ & WSe$_{2}$ \\ \hline
A & Mass/$m_{0}$ & 1.1 & 1.33 & 0.84 & 0.93 \\
& $D_{0},$ {\AA} & 0.29 & 0.23 & 0.32 & 0.30 \\ \hline
B & Mass/$m_{0}$ & 0.98 & 1.15 & 0.62 & 0.66 \\
& $D_{0},$ {\AA} & 0.31 & 0.25 & 0.37 & 0.36 \\ \hline\hline
\end{tabular}%
\end{center}
\label{tab6}
\end{table}

The binding energy of  an exciton depends on the effective masses of
 an electron and  a hole that constitute the A and B
excitons and  the polarizability of TMDC material. In our
calculations throughout of this paper we use the set of effective
masses for electrons and holes and  the polarizability from
Refs.~\cite{emass,hmass}. The effective masses for the charge
carriers in Refs.~\cite{emass,hmass} and the polarizability in Ref.
\protect\cite{hmass} were obtained by employing density functional
theory (DFT) calculations.

The binding energies for A and B dipolar excitons in
different TMDC double layers, formed by the same monolayers, are
represented in Table~\ref{tab3}. The highest binding energy
corresponds to a MoSe$_{2}$ double layer, while the lowest
corresponds to  a WS$_{2}$ double layer.

\begin{table}[t]
\caption{Binding energy in eV for A and B type dipolar excitons in
 a TMDC double layer with the same  TMDC
materials  }
\begin{center}
\begin{tabular}{c|cccc|cccc}
\hline\hline $N_{L}$ h-BN & \multicolumn{4}{|c|}{Binding energy of A
exiton, eV} & \multicolumn{4}{|c}{Binding energy of B exiton, eV} \\
\cline{2-9}
& MoS$_{2}$ & MoSe$_{2}$ & WS$_{2}$ & WSe$_{2}$ & MoS$_{2}$ & MoSe$_{2}$ & WS%
$_{2}$ & WSe$_{2}$ \\ \hline
8 & 0.049 & 0.054 & 0.040 & 0.041 & 0.045 & 0.050 & 0.029 & 0.030 \\
10 & 0.045 & 0.049 & 0.038 & 0.040 & 0.043 & 0.046 & 0.028 & 0.029 \\
12 & 0.042 & 0.044 & 0.036 & 0.038 & 0.039 & 0.042 & 0.028 & 0.029 \\
\hline\hline
\end{tabular}
\end{center}
\label{tab3}
\end{table}

For the  double layer heterostructure, formed by two different TNDC
monolayers,  if the transition metal atom is replaced with another
transition metal atom  (e.g. for a MoS$_2$/WS$_{2}$ heterostructure)
or if the chalcogenide atoms are replaced with the other
chalcogenide atoms (e.g. for a MoS$_{2}$/MoSe$_{2}$
heterostructure), we estimated the polarizability as the average of
the polarizabilities of two TMDC monolayers. The binding energies
for A and B dipolar excitons in different TMDC double layers, formed
by  two different monolayers, are represented in Tables~\ref{tab4}
and~~\ref{tab5}, respectively. The highest binding energy
corresponds to a double layer with electrons in a MoSe$_{2}$
monolayer and holes in a MoS$_{2}$ monolayer.

Our calculation of the dipolar exciton binding energy by employing
the harmonic oscillator approximation for the Keldysh potential for
electron-hole attraction shows that the binding energy for A
excitons is greater than for B excitons for the TMDC double layer
heterostructure with the same monolayers. The latter is related to
the fact that the effective masses of  spin-down holes from the
valence band are sufficiently larger than the effective masses of
spin-up holes due to the strong spin-orbit coupling on the valence
band of TMDCs, and
 the effective  masses of spin-up electrons from the conduction band  are slightly
greater than the effective masses of spin-down electrons. Therefore,
the reduced mass $\mu$ is larger for A excitons than for B excitons,
which results in higher binding energies for A excitons than for B
excitons. Let us mention that it is possible to create either
electrons or holes in either TMDC monolayer by changing the
direction of the electric field perpendicular to the plane of the
double layer. Since the effective masses of electrons and holes are
different for the same TMDC monolayers, the binding energy for
dipolar excitons depends on a choice of a monolayer for electrons
and holes, correspondingly. According
to Tables~\ref{tab4} and~~\ref{tab5}, the binding energies for
dipolar excitons with electrons in MoS$_{2}$ and MoSe$_{2}$ and
holes in WS$_{2}$ and WSe$_{2}$ are larger than for dipolar excitons
with electrons in WS$_{2}$ and WSe$_{2}$ and holes in MoS$_{2}$  and
MoSe$_{2}$.

\begin{table}[b]
\caption{Binding energy in eV for A type dipolar exciton in
 a  TMDC double layer with different TMDC materials}
\label{tabAenergy}
\begin{center}
\begin{tabular}{c|c|ccc|ccc|ccc|ccc}
\hline\hline
$N_{L}$ h-BN & Electron layer & \multicolumn{3}{|c|}{MoS$_{2}$} &
\multicolumn{3}{|c|}{MoSe$_{2}$} & \multicolumn{3}{c|}{WS$_{2}$} &
\multicolumn{3}{c}{WSe$_{2}$} \\ \cline{2-14}
& Hole layer & MoSe$_{2}$ & \multicolumn{1}{|c}{WS$_{2}$} &
\multicolumn{1}{|c|}{WSe$_{2}$} & MoS$_{2}$ & \multicolumn{1}{|c}{WS$_{2}$}
& \multicolumn{1}{|c|}{WSe$_{2}$} & MoS$_{2}$ & \multicolumn{1}{|c}{MoSe$_{2}
$} & \multicolumn{1}{|c|}{WSe$_{2}$} & MoS$_{2}$ & \multicolumn{1}{|c}{MoSe$%
_{2}$} & \multicolumn{1}{|c}{WS$_{2}$} \\ \hline
8 & $B,$ eV & 0.050 & 0.045 & 0.047 & 0.053 & 0.049 & 0.050 &
0.042 & 0.043
& 0.045 & 0.042 & 0.034 & 0.046 \\
10 & $B,$ eV & 0.046 & 0.043 & 0.044 & 0.048 & 0.045 & 0.046 & 0.041
& 0.041
& 0.043 & 0.039 & 0.033 & 0.044 \\
12 & $B,$ eV & 0.042 & 0.039 & 0.040 & 0.044 & 0.042 & 0.042 & 0.038
& 0.038 & 0.041 & 0.036 & 0.032 & 0.042 \\ \hline\hline
\end{tabular}%
\label{tab4}
\end{center}
\end{table}

\begin{table}[t]
\caption{Binding energy in eV for B type dipolar exciton in
 a TMDC double layer with different TMDC materials}
\begin{center}
\begin{tabular}{c|c|ccc|ccc|ccc|ccc}
\hline\hline
$N_{L}$ h-BN & Electron layer & \multicolumn{3}{|c|}{MoS$_{2}$} &
\multicolumn{3}{|c|}{MoSe$_{2}$} & \multicolumn{3}{c|}{WS$_{2}$} &
\multicolumn{3}{c}{WSe$_{2}$} \\ \cline{2-14}
& Hole layer & MoSe$_{2}$ & \multicolumn{1}{|c}{WS$_{2}$} &
\multicolumn{1}{|c|}{WSe$_{2}$} & MoS$_{2}$ & \multicolumn{1}{|c}{WS$_{2}$}
& \multicolumn{1}{|c|}{WSe$_{2}$} & MoS$_{2}$ & \multicolumn{1}{|c}{MoSe$_{2}
$} & \multicolumn{1}{|c|}{WSe$_{2}$} & MoS$_{2}$ & \multicolumn{1}{|c}{MoSe$%
_{2}$} & \multicolumn{1}{|c}{WS$_{2}$} \\ \hline
8 & $B,$ eV & 0.046 & 0.038 & 0.038 & 0.049 & 0.041 & 0.041 &
0.035 & 0.035
& 0.031 & 0.034 & 0.027 & 0.034 \\
10 & $B,$ eV & 0.043 & 0.037 & 0.037 & 0.045 & 0.039 & 0.039 & 0.034
& 0.035
& 0.033 & 0.033 & 0.027 & 0.035 \\
12 & $B,$ eV & 0.040 & 0.035 & 0.035 & 0.041 & 0.037 & 0.037 & 0.033
& 0.034 & 0.032 & 0.032 & 0.026 & 0.034 \\ \hline\hline
\end{tabular}%
\label{tab5}
\end{center}
\end{table}


The energy spectrum of the center-of-mass of the A (or B) dipolar
exciton  $\varepsilon_{0}^{A(B)}(\mathbf{P})$   is given by
\begin{eqnarray}
\varepsilon_{0}^{A(B)}(\mathbf{P}) = \frac{P^{2}}{2M_{A(B)}},
\label{eps0}
\end{eqnarray}
where $P$ is a momentum of the center of mass of a dipolar exciton,
and the masses of A and B dipolar excitons are given by $M_{A} =
m_{e\uparrow} + m_{h\downarrow}$ and $M_{B} = m_{e\downarrow} +
m_{h\uparrow}$, where $m_{e\uparrow(\downarrow)}$ is the effective
mass of spin-up (spin-down) electrons from the conduction band and
$m_{h\uparrow(\downarrow)}$ is the effective mass of spin-up
(spin-down) holes from the valence band, correspondingly.

\section{Exciton-exciton interaction}

\label{exexint}

We consider a dilute system  of electrons and holes in two  parallel
TMDC monolayers, spatially separated by a dielectric, when $n
r_{X}^{2}\ll 1$, where $n$  is the  2D concentration for dipolar
excitons. In this case, the dipolar excitons are formed by
electron-hole pairs with the electrons and holes spatially separated
in two different TMDC monolayers. One can estimate the concentration
of excitons $n$ that corresponds to   $r_{X}^{-2}$  by using
Eq.~(\ref{rx2}). $r_{X}^{2}$ depends on the interlayer separation
and when $D$ increases $r_{X}^{2}$ increases. For all TMDC materials
$r_{X}^{2}$ varies from $7.4\times 10^{-18}$ cm$^{2}$ to $9.1\times
10^{-18}$ cm$^{2}$ that corresponds to the  2D concentrations
$1.4\times 10^{17}$ cm$^{-2}$ and $1.1\times 10^{17}$ cm$^{-2}$ when
the interlayer separation is $N_L=12$ of $h$-BN monolayers. For the
smaller interlayer separation the corresponding concentration even
is larger. Below we are considering the dilute limit with
 the maximal 2D exciton concentration that not exceeded
$6\times10^{15}$ cm$^{-2}$, where the model of a weakly interacting
Bose gas is valid.

The  excitons, which are not   elementary but composite
bosons~\cite{Combescot} are different from bosons only due to
exchange effects~\cite{Moskalenko_Snoke}. At large interlayer
separations $D$,  the exchange effects in the exciton-exciton
interactions in a TMDC double layer are negligible, because the
exchange interactions in a system of spatially separated electrons
and holes in a double layer are suppressed due to the low tunneling
probability, caused by  shielding of the dipole-dipole interaction
by the insulating barrier~\cite{BK}. Therefore,  the dilute system
of dipolar excitons in a TMDC double layer can be treated as a
weakly interacting Bose gas.

Assuming that at $T=0$ K almost all dipolar excitons belong to a
BEC,  this two-component weakly interacting Bose gas of A and B
dipolar excitons can be treated in the framework of the Bogoliubov
approximation~\cite{Abrikosov,Lifshitz}. Within the Bogoliubov
approximation for a weakly interacting Bose gas, the many-particle
Hamiltonian can be  diagonalized,   replacing the product of four
operators in the interaction term by the product of two operators.
This is valid if most of the particles belong to the BEC, and only
the interactions between the condensate and non-condensate particles
are considered, while the interactions between non-condensate
particles are neglected. The operators for condensate bosons are
replaced by numbers \cite{Abrikosov}, and the resulting Hamiltonian
is quadratic with respect to the creation and annihilation
operators.

Let us consider and present the expressions for the coupling
constant $g$ for the both Keldysh and Coulomb potentials for
interaction between the charge carriers. When two dipolar excitons
are separated by the distance $R$ and the electron and hole of one
dipolar exciton interact with the electron and hole of another
dipolar exciton, then the exciton-exciton interaction potential
$U(R)$ can be presented as
\begin{equation}
U(R)=2V(R)-2V\left(R\sqrt{1+\frac{D^{2}}{R^{2}}}\right),
\label{Keldysh Dipole}
\end{equation}%
where $V(R)$ is  the potential for interaction between two electrons
or two holes in the same TMDC monolayer. The interaction $V(R)$ can
 be given either by the Keldysh potential
(\ref{Keldysh}) or by the Coulomb potential.

In a very dilute weakly-interacting Bose gas of dipolar excitons,
implying  $D \ll R$, the second term in Eq. (\ref{Keldysh Dipole})
can be expanded in terms of $(D/R)^{2},$ and by retaining only the
first order terms with respect to $(D/R)^{2}$, one gets

\begin{eqnarray}   \label{Dipolar Keldysh Approx}
U(R)=\left\{
\begin{array}{c}
\frac{\pi ke^{2}D^{2}}{2\varepsilon _{d}\rho _{0}^{2}R}\left[
Y_{-1}\left(
\frac{R}{\rho _{0}}\right) -H_{-1}(y)\left( \frac{R}{\rho _{0}}\right) %
\right],\text{ for the Keldysh potential, } \\
\frac{ke^{2}D^{2}}{\varepsilon _{d}R^{3}},\text{ \ \ \ \ \ \ \ \ \ \
\ \ \ \ \ \ \ \ \ \ \ \ \ \ \ \ \ \ \ \ \ \ \ \ \ \ \ \   for the
Coulomb potential.}%
\end{array}%
\right.
\end{eqnarray}

\medskip
\par

According to the procedure presented in Ref.\
[\onlinecite{BK,BKKL}], the exciton-exciton coupling constant $g$ in
a very dilute Bose gas of A and B excitons can be obtained under the
assumption that  the dipole-dipole repulsion of dipolar excitons
 exists only at distances
between excitons larger than distance from the exciton to the
classical turning point. The separation between two dipolar excitons
cannot be smaller than this distance, and the coupling constants for
the exciton-exciton interaction is obtained as
\begin{eqnarray}
g_{i} =   2\pi \int_{R_{0i}}^{\infty} R dR\ U(R) , \ \ i=AA,\ BB, \
AB  , \label{gK1}
\end{eqnarray}
where $R_{0AA}$, $R_{0BB}$, and $R_{0AB}$ are the distances between
two dipolar excitons at the classical turning point for two A
excitons, two B excitons, and one A and one B excitons,
correspondingly.

Substituting Eq.~~(\ref{Dipolar Keldysh Approx}) into
Eq.~~(\ref{gK1}),  the exciton-exciton coupling constant $g$ can be
written as following
\begin{eqnarray}  \label{gK}
g_{i}=\left\{
\begin{array}{c}
\frac{\pi ^{2}ke^{2}D^{2}}{\varepsilon _{d}\rho _{0}}\left[ H_{0}\left( \frac{%
R_{0 i}}{\rho _{0}}\right) -Y_{0}\left( \frac{R_{0 i}}{\rho
_{0}}\right) \right]
,\text{ for the Keldysh potential,} \\
\frac{2\pi ke^{2}D^{2}}{\varepsilon _{d}R_{0i}},\text{ \ \ \ \ \ \ \
\ \ \ \ \ \
\ \ \ \ \ \ \ \ \ \ \ \ \ \ \ \ \ \ \  \  for the Coulomb potential.%
}%
\end{array}%
\right.
\end{eqnarray}

The system of equations for $R_{0AA}$, $R_{0BB}$, and $R_{0AB}$ is
derived in Appendix~\ref{app:A}. The system of
equations~(\ref{geq1res}) has all real and positive roots only if
$y_{AA}=y_{BB}=y_{AB}\equiv y$, which implies
$R_{0AA}=R_{0BB}=R_{0AB}\equiv R_{0}$ and
$g_{AA}=g_{BB}=g_{AB}\equiv g$. In a very dilute system, the
exciton-exciton coupling constant  is the same for all three
possible combinations of A and B excitons, because at large
distances within the dipole approximation~(\ref{Keldysh Dipole}) the
exciton-exciton interaction potential $U(R)$, given by
Eq.~(\ref{Dipolar Keldysh Approx}), depends only on the charges of
electrons and holes, which are the same for A and B dipolar
excitons. Combining Eqs.~(\ref{geq1}),~(\ref{Dipolar Keldysh Approx})
and~(\ref{gK}), for the Keldysh potential one obtains the following transcendental
equation for $R_{0}$:
\begin{eqnarray}
4 \pi n \rho_{0}^{2} y \left[ H_{0}(y)-Y_{0}(y)\right] = - \left[
H_{-1}(y)-Y_{-1}(y)\right]
  , \label{R0K}
\end{eqnarray}
where $y = R_{0}/\rho_{0}$.

For the Coulomb potential one  combines
Eqs.~~(\ref{geq1}),~(\ref{Dipolar Keldysh Approx}),~(\ref{gK}), implies $R_{0AA}=R_{0BB}=R_{0AB}\equiv R_{0}$

and derives the following expression for $R_{0}$:

\begin{eqnarray}
R_{0} =  \frac{1}{2\sqrt{\pi n}}  . \label{r0}
\end{eqnarray}

 From Eqs.\ (\ref{r0}),~(\ref{gK}) and~(\ref{mu1}), we derive  the
exciton-exciton coupling constant $g$ for the Coulomb potential

\begin{eqnarray}
g=\frac{4\pi ke^{2}D^{2}\sqrt{\pi n}}{\varepsilon _{d}} .
\label{geqeq1}
\end{eqnarray}

\begin{figure}[!b]
\centering
\includegraphics[width=12.0cm]{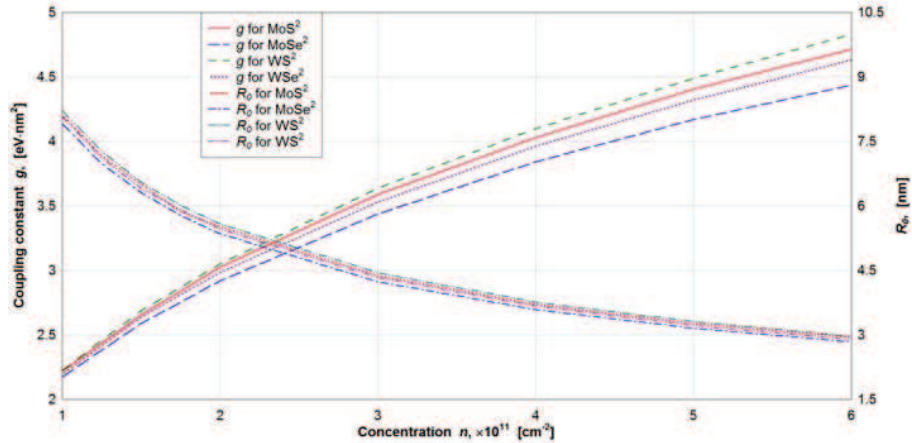}
\caption{The coupling constant $g$ and the distance $R_{0}$ between
two dipolar excitons at the classical turning point for different
TMDC double layers as functions of the exciton concentration. The
number of h-BN monolayers between the TMDC monolayers is $N_{L} =
10$. The calculations were performed for  the polarizabilities from
Ref.~\cite{hmass}. } \label{gfig}
\end{figure}

From Eq.\ (\ref{R0K}) it follows that for the Keldysh potential
$R_{0}$ depends on the exciton concentration and the type of TMDC
material in the double layer, while for the Coulomb potential
$R_{0}$ depends only on the excitonic concentration according to
Eq.~(\ref{r0}). The corresponding dependence is replicated for the
exciton-exciton coupling constant \textit{g}, while the coupling
constant has the same dependence on the interlayer separation for
the both potentials and is directly proportional  to $D^{2}$. The
coupling constant $g$ and the distance $R_{0}$ between two dipolar
excitons at the classical turning point for different TMDC double
layers as functions of the exciton concentration $n$ for the Keldysh
potential are represented in Fig.~\ref{gfig}.
According to Fig.~\ref{gfig}, $g$ increases and $R_{0}$ decreases as
the exciton concentration $n$ increases. While the values of $g$ and
$R_{0}$ are close to each other for different TMDC double layers,
the largest $g$ and $R_{0}$ correspond to a WS$_{2}$ double layer,
and the smallest $g$ and $R_{0}$ correspond to a MoSe$_{2}$ double
layer. According to Fig.~\ref{gfig}, the difference between
exciton-exciton coupling constants $g$ for different TMDC double
layers increases as the exciton concentration $n$ increases. Since
for the Keldysh potential  $g$ and $R_{0}$ depend on the TMDC
material of the double layer only through the polarizability of the
material and do not depend on the electrons and holes effective
masses, the exciton-exciton coupling constant  $g$ does not depend
on the choice of the monolayer for  either electrons or holes, where
the transition metal atom is replaced with the other one and/or  the
chalcogenide atoms are replaced with the other ones. We did not
display $g$ and $R_{0}$ for TMDC double layers, formed by the
monolayers of different materials, because the results are very
close to ones  for two monolayers of the same TMDC material,
presented in Fig.~\ref{gfig}.

The spectrum of collective excitations for a two-component weakly
interacting Bose gas of A and B excitons was derived within the
Bogoliubov  approximation~\cite{BK}. In the limit of small momenta
$P$, if the densities of A and B excitons are the same and
$n_{A}=n_{B}=n/2$, the spectrum of collective excitations is
$\varepsilon(P) = c_{S}P$, where $c_{S}$ is the sound velocity. The latter one is  given by~\cite{BK}
\begin{eqnarray}  \label{c2}
c_{S}=\sqrt{\frac{gn}{2}\left(
\frac{1}{M_{A}}+\frac{1}{M_{B}}\right) },
\end{eqnarray}
where $g$ is a coupling constant for the interaction between two
dipolar excitons, defined above by Eq. (\ref{gK}).

\section{Superfluidity}

\label{sup}

The weakly interacting Bose gas of dipolar excitons in a TMDC double
layer satisfies to the Landau criterion for superfluidity
\cite{Abrikosov,Lifshitz}, because  at small momenta, the energy
spectrum of the quasiparticles is sound-like. The critical exciton
velocity for superfluidity is given by $v_{c} = c_{S}$, because the
quasiparticles can be created only at velocities above the sound
velocity.

 Within the mean field approximation, the superfluidity
in the dilute system of dipolar excitons in a TMDC double layer
occurs at the temperatures below the mean field critical temperature
of superfluidity $T_{c}$, which  is given by~\cite{BK}
\begin{eqnarray}  \label{Tc2e}
T_{c}= \frac{1}{k_{B}}  \left[ \frac{\pi \hbar
^{2}g^{2}n^{3}}{12\zeta (3)}Q\right] ^{1/3}  .
\end{eqnarray}
where $k_{B}$ is the Boltzmann constant, $\zeta (z)$ is the Riemann zeta function ($\zeta
(3)\simeq 1.202$), and the factor $Q$ is defined as
\begin{eqnarray}  \label{Qdef}
Q=\frac{M_{A}+M_{B}}{\left( \mu _{AB}\right)^{2}}  .
\end{eqnarray}
In Eq.~(\ref{Qdef})  $\mu _{AB}=M_{A}M_{B}/(M_{A}+M_{B})$ is the reduced mass for
two-component system of A and B excitons, $M_{A}$ and $M_{B}$ are
the effective masses of A and B excitons, correspondingly.

\begin{figure}[h]
\centering
\includegraphics[width=12.0cm]{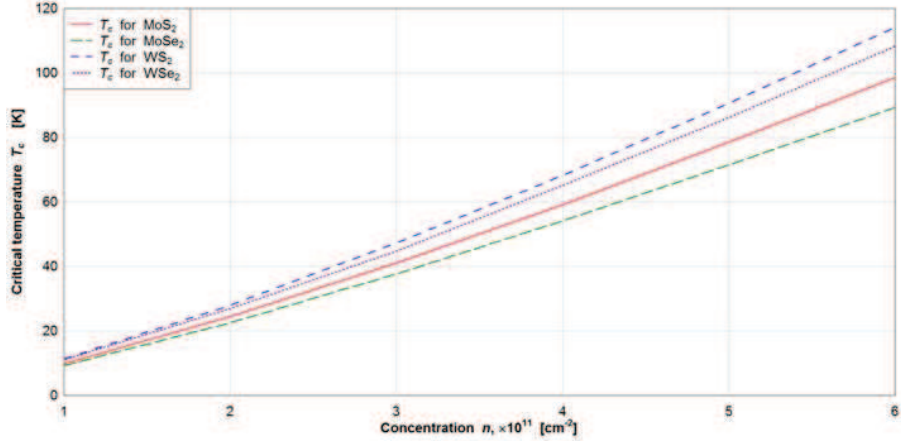}
\caption{The mean field critical temperature of the superfluidity
$T_{c}$  as a function of the exciton concentration $n$ for
different TMDC double layers. The calculations were performed for
the number $N_{L} = 10$ of $h$-BN monolayers, located between two
TMDC monolayers,  The calculations were performed for the
polarizabilities and the  set of effective masses for electrons and
holes  from Refs. \protect\cite{emass,hmass}. } \label{Tcdif}
\end{figure}

The mean field critical temperature of the superfluidity $T_{c}$ for
the dipolar excitons formed via the Keldysh potential as a function
of the exciton concentration $n$ for different TMDC double layers is
shown in Fig.~\ref{Tcdif}. According to Fig.~\ref{Tcdif}, $T_{c}$
increases as $n$ increases. The largest $T_{c}$ corresponds to a
$\mathrm{W S_{2}}$ double layer, while the smallest $T_{c}$
corresponds to a $\mathrm{Mo Se_{2}}$ double layer, that correlated
with the corresponding $Q$ factor presented in Table~\ref{tab1}. Of
course, the mean field critical temperature depends on the
polarizability of TMDC material. However, the $Q$ factor has the
major impact. Interestingly enough, the corresponding binding energy
for the $\mathrm{W S_{2}}$ is smaller than for the $\mathrm{Mo
Se_{2}}$. In  Fig.~\ref{Tcdif}, we show $T_{c}$ for the number
$N_{L} = 10$ of $h$-BN monolayers, located between two TMDC
monolayers. The increase of  $D$ results in  the increase of the
potential barrier for electron-hole tunneling between the layers,
which leads to  the increase of the exciton lifetime. Besides,
larger $D$ results in the increase of the exciton dipole moment and
the exciton-exciton coupling constant $g$. Therefore, according to
Eq.~(\ref{c2}) the sound velocity increases with the increase of
$g$, which  leads to the increase of the mean field critical
temperature of superfluidity with the increase of $D$. As it is
demonstrated  in Fig.~\ref{Fig2}, $T_{c}$ increases when $D$
increases.

\begin{figure}[h]
\centering
\includegraphics[width=12.0cm]{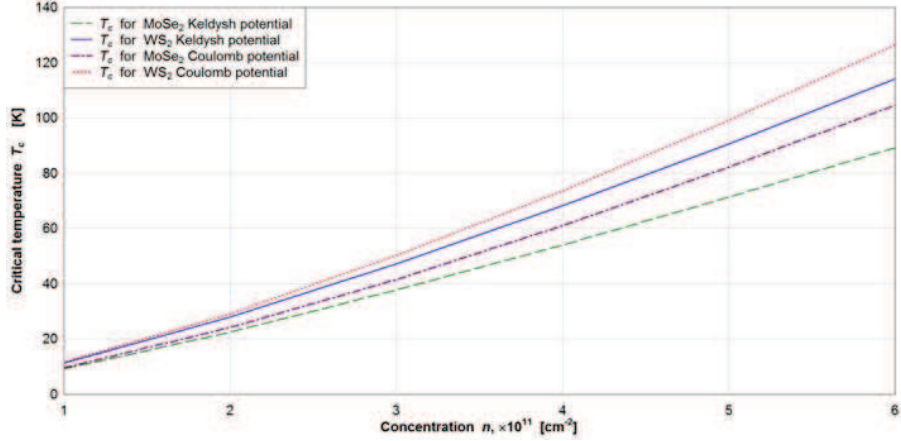}
\caption{The mean field critical temperature of the superfluidity
$T_{c}$  as a function of the exciton concentration $n$ for
$\mathrm{W S_{2}}$ and $\mathrm{Mo Se_{2}}$ double layers for the
Keldysh and Coulomb potentials of interaction between the charge
carriers. The calculations were performed for the number $N_{L} =
10$ of $h$-BN monolayers, located between two TMDC monolayers. The
calculations were performed for the polarizabilities and  the set of
effective masses for electrons and holes from Refs.
\protect\cite{emass,hmass}. } \label{TcKC}
\end{figure}

\begin{figure}[h]
\centering
\includegraphics[width=12.0cm]{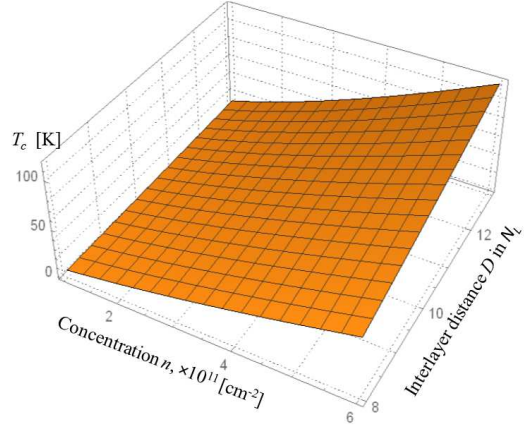}
\caption{(Color online) The mean field critical temperature of the
superfluidity $T_{c}$ for a $\mathrm{W S_{2}}$ double layer as a
function of the exciton concentration $n$ and the interlayer
separation $D$. The calculations were performed  for the set of
effective masses for electrons and holes from Refs.
\protect\cite{emass,hmass}. }
 \label{Fig2}
\end{figure}

Let's compare the critical temperatures obtained for the Keldysh and Coulomb potentials. The mean field critical temperature of the superfluidity $T_{c}$  as
a function of the exciton concentration $n$ for $\mathrm{W S_{2}}$
and $\mathrm{Mo Se_{2}}$ double layers for the Keldysh and Coulomb
potentials of interaction between the charge carriers is represented
in Fig.~\ref{TcKC}. According to Fig.~\ref{TcKC}, $T_{c}$ for the
Keldysh potential is smaller than for the Coulomb potential due to
the screening effects, taken into account by the Keldysh potential.
The screening effects cause the exciton-exciton interaction to
become weaker, which leads to  smaller exciton-exciton coupling
constant $g$, resulting in smaller $T_{c}$.

It is important to mention that while for the Coulomb potential
$T_{c}$ depends on the TMDC materials, forming a double layer, only
through the effective exciton masses, constituting  the factor $Q$, for
the Keldysh potential $T_{c}$ depends on the TMDC materials  also
through the exciton-exciton coupling constant $g$, determined by the
polarizability of the TMDC materials.

The mean field critical temperature of the
superfluidity $T_{c}$ for
 a $\mathrm{W S_{2}}$ double layer under the assumption about the Coulomb  potential
for the interaction between the  charge carriers  is represented as
a function of the exciton concentration $n$ and the interlayer
separation $D$ in Fig.~\ref{Fig2}. We choose to plot $T_{c}$ for  a
$\mathrm{WS_{2}}$ double layer, since $T_{c}$  for this material is
larger than for other materials at the same $n$ and $D$. According
to Fig.~\ref{Fig2}, $T_{c}$ increases as $n$ and $D$ increase.

 Let us mention that while for a one-component Bose gas
the critical temperature of superfluidity is a decreasing function
of the mass of a particle, for a two-component weakly interacting
Bose gas of A and B dipolar excitons $T_{c}$ is an increasing
function of the factor $Q$ according to Eq.~(\ref{Tc2e}), where the
dependence of the factor $Q$ on the effective masses of A and B
dipolar excitons is given by Eq.~(\ref{Qdef}). The factors $Q$ for A
and B type excitons in different TMDC double layers, formed by two
monolayers of the same material, are presented in Table~\ref{tab1}.
 The factors $Q$ for A
and B type excitons in different TMDC double layers, formed by two
monolayers of two different materials, when the transition metal
atom is replaced by the other transition metal atom (e.g. for a
MoS$_2$/WS$_{2}$ heterostructure)  or when  the chalcogenide atoms
are replaced by the other chalcogenide atoms (e.g. for a
MoS$_{2}$/MoSe$_{2}$ heterostructure), are presented in
Table~\ref{tab2}. According to Tables~\ref{tab1} and~\ref{tab2}, the
largest factor $Q$ corresponds to a double layer, formed by two
WS$_{2}$ monolayers, while the smallest factor $Q$ corresponds to a
double layer, formed by two MoSe$_{2}$ monolayers. The critical
temperature of superfluidity $T_c$  in different combinations of
TMDC double layers for different densities   and corresponding $Q$
factor are shown in Table~\ref{tabTemperature}.  Since the effective
masses of electrons and holes are different for the same TMDC
monolayers, the factor $Q$ and $T_{c}$ depend on a choice of a
monolayer for electrons and holes, correspondingly. According to
Table~\ref{tabTemperature}, for a given TMDC double layer there is a
correlation between $Q$ and $T_{c}$, implying that higher $T_{c}$
corresponds to the double layer with higher $Q$, as follows from
Eq.~(\ref{Tc2e}). As it can be seen in  Table~\ref{tabTemperature},
the largest $T_{c}$ and $Q$ correspond to a double layer, formed by
two WS$_{2}$ monolayers, while the smallest $T_{c}$ and $Q$
correspond to a double layer, formed by two MoSe$_{2}$ monolayers.

\begin{table}[t]
\caption{$Q$ factor in units of $1/m_{0}$, reduced mass $\protect\mu
_{AB}$ and $M_{A}+M_{B}$ in units of $m_{0}$ for A and B type
dipolar excitons in  a TMDC double layer with the same TMDC
materials}
\begin{center}
\begin{tabular}{ccccc}
\hline\hline & MoS$_{2}$ & MoSe$_{2}$ & WS$_{2}$ & WSe$_{2}$ \\
\hline
$Q,$ $[1/m_{0}]$ & 7.74 & 6.52 & 11.47 & 10.67 \\
$\mu _{AB},$ $[m_{0}]$ & 0.52 & 0.62 & 0.36 & 0.39 \\
$M_{A}+M_{B},$ $[m_{0}]$ & 2.08 & 2.48 & 1.46 & 1.49 \\ \hline\hline
\end{tabular}
\end{center}
\label{tab1}
\end{table}

\begin{table}[b]
\caption{$Q$ factor in units of $1/m_{0}$, reduced mass $\protect\mu
_{AB}$ and $M_{A}+M_{B}$ in units of $m_{0}$ for A and B type
dipolar excitons in   a TMDC double layer with different TMDC
materials}
\begin{center}
\begin{tabular}{c|ccc|ccc|ccc|ccc}
\hline\hline
Electron layer & \multicolumn{3}{|c|}{MoS$_{2}$} & \multicolumn{3}{|c|}{MoSe$%
_{2}$} & \multicolumn{3}{c|}{WS$_{2}$} & \multicolumn{3}{c}{WSe$_{2}$} \\
\hline
Hole layer & MoSe$_{2}$ & \multicolumn{1}{|c}{WS$_{2}$} &
\multicolumn{1}{|c|}{WSe$_{2}$} & MoS$_{2}$ & \multicolumn{1}{|c}{WS$_{2}$}
& \multicolumn{1}{|c|}{WSe$_{2}$} & MoS$_{2}$ & \multicolumn{1}{|c}{MoSe$_{2}
$} & \multicolumn{1}{|c|}{WSe$_{2}$} & MoS$_{2}$ & \multicolumn{1}{|c}{MoSe$%
_{2}$} & \multicolumn{1}{|c}{WS$_{2}$} \\ \hline
$Q,$ $[1/m_{0}]$ & 7.31 & 9.25 & 9.05 & 6.86 & 8.03 & 7.88 &
9.17 & 8.57 &
11.2 & 8.80 & 8.25 & 10.9 \\
$\mu _{AB},$ $[m_{0}]$ & 0.55 & 0.44 & 0.45 & 0.59 & 0.50 & 0.52 &
0.44 &
0.47 & 0.37 & 0.46 & 0.49 & 0.38 \\
$M_{A}+M_{B},$ $[m_{0}]$ & 2.21 & 1.77 & 1.82 & 2.35 & 2.04 & 2.09 &
1.77 & 1.90 & 1.51 & 1.85 & 1.98 & 1.54 \\ \hline\hline
\end{tabular}
\end{center}
\label{tab2}
\end{table}

\begin{table}[t]
\caption{Critical temperature of superfluidity $T_c$ in K in
different combinations of TMDC double layers for diffident densities
and  the number $N_{L} = 10$ of $h$-BN monolayers, located between
two TMDC monolayers,  and corresponding $Q$ factor in units of
$1/m_{0}$}
\begin{center}
\begin{tabular}{c|cccc|cccc|cccc|cccc}
\hline\hline
Electron layer & \multicolumn{4}{|c|}{MoS$_{2}$} & \multicolumn{4}{|c|}{MoSe$%
_{2}$} & \multicolumn{4}{c|}{WS$_{2}$} & \multicolumn{4}{c}{WSe$_{2}$} \\
\hline
Hole layer & MoS$_{2}$ & \multicolumn{1}{|c}{MoSe$_{2}$} &
\multicolumn{1}{|c}{WS$_{2}$} & \multicolumn{1}{|c|}{WSe$_{2}$} & MoS$_{2}$
& \multicolumn{1}{|c}{MoSe$_{2}$} & \multicolumn{1}{|c}{WS$_{2}$} &
\multicolumn{1}{|c|}{WSe$_{2}$} & MoS$_{2}$ & \multicolumn{1}{|c}{MoSe$_{2}$}
& \multicolumn{1}{|c}{WS$_{2}$} & \multicolumn{1}{|c|}{WSe$_{2}$} & MoS$_{2}$
& \multicolumn{1}{|c}{MoSe$_{2}$} & \multicolumn{1}{|c}{WS$_{2}$} &
\multicolumn{1}{|c}{WSe$_{2}$} \\ \hline
$n=3\times 10^{11}$ cm$^{-2}$ & 41 & 39 & 44 & 43 & 40 & 38 & 41 &
40 & 43 &
42 & 47 & 47 & 43 & 41 & 46 & 45 \\
$n=5\times 10^{11}$ cm$^{-2}$ & 79 & 76 & 84 & 82 & 77 & 71 & 79 &
77 & 83 &
80 & 91 & 89 & 81 & 78 & 88 & 86 \\
$Q,$ $[1/m_{0}]$ & 7.74 & 7.31 & 9.25 & 9.05 & 7.54 & 6.52 & 8.03 & 7.88 &
9.17 & 8.57 & 11.47 & 11.2 & 8.80 & 8.25 & 10.9 & 10.67 \\ \hline\hline
\end{tabular}%
\end{center}
\label{tabTemperature}
\end{table}

\section{Conclusions}

\label{conc}

 In this paper we have studied the formation and superfluidity of dipolar excitons in double layer
 heterostructures that  are formed by two monolayers of the same TMDC material and two different TMDC monolayers.
 In the framework of the harmonic oscillator approximation for the Keldysh potential the analytical expressions for
 the exciton energy spectrum and the mean field critical temperature of superfluidity are obtained. All calculations are performed
 by using the effective electron and hole masses and polarizability obtained within the framework of
 DFT approach for the TMDC material. We have calculated the binding energies for A and B dipolar excitons in
 various TMDC double layers, taking into account screening effects by employing the approximated
 Keldysh potential for the interaction between the charge carriers. The binding energy of dipolar
 excitons depends on  the electron and  hole reduced mass,  the polarizability of the TMDC material and  the interlayer
 separation  between two monolayers.  Since
the effective masses of electrons and holes are different for the
same TMDC monolayers,  the exciton binding energy is larger for
dipolar excitons with electrons in MoS$_{2}$ and MoSe$_{2}$ and
holes in WS$_{2}$ and WSe$_{2}$ than   for dipolar excitons with
electrons in WS$_{2}$ and WSe$_{2}$ and holes in MoS$_{2}$ and
MoSe$_{2}$.  The increase of the electron and hole reduced mass
leads to  the decrease of the binding energy of dipolar exciton. The
mean field critical temperature of superfluidity $T_{c}$ for a
dilute system of A and B dipolar excitons in TMDC materials are
determined by the exciton-exciton coupling constant $g$ and the
factor $Q$, which depends of the sum of the effective masses of A
and B excitons and the reduced mass for  a two-component system of A
and B excitons. Different effective electron and hole masses result
in different effective masses of A and B excitons $M_{A}$ and
$M_{B}$, a different reduced mass $\mu _{AB}$ for two-component
system of A and B excitons, and a different factor $Q$. We have
calculated the exciton-exciton coupling constant $g$ and the mean
field critical temperature superfluidity $T_{c}$ for a dilute
two-component  system of  A and B dipolar excitons. Let us mention
that $T_{c}$ for a two-component weakly interacting gas of A and B
dipolar excitons is an increasing function of the factor $Q$,
determined by the effective reduced mass of A and B excitons, which
is always smaller than the individual effective mass of either A or
B exciton. While the factor $Q$ and $T_{c}$ depend on the exciton
effective masses, the exciton-exciton coupling constant $g$ and the
distance between two dipolar excitons at the classical turning point
$R_{0}$ do not depend on the exciton effective masses but depend on
the exciton concentration and  the polarizability of TMDC materials.
The critical temperature of superfluidity  $T_{c}$, calculated by
using the harmonic potential approximation of the Keldysh potential
for the interaction between the charge carriers, is smaller than for
the Coulomb potential due to diminishing exciton-exciton interaction
by screening effects, taken into account by the Keldysh potential.
These screening effects lead to the decrease of the exciton-exciton
coupling constant, which results in the decrease of $T_{c}$. The
largest mean field critical temperature $T_{c}$ of two-component
superfluidity  of A and B dipolar excitons was obtained for a
$\mathrm{W S_{2}}$ double layer, while the smallest $T_{c}$ was
obtained for a $\mathrm{Mo Se_{2}}$ double layer. For a given 2D
exciton concentration $n$ and interlayer separation $D$, the TMDC
double layer with higher $T_{c}$ corresponds to the double layer
with higher $Q$.  By comparison of the exciton-exciton coupling $g$
and the mean field critical temperature $T_{c}$ of superfluidity
calculated by using the Keldysh and Coulomb potential for the
interaction between the charge carriers, one can study the influence
of the screening effects on  a weakly interacting gas of dipolar
excitons in a TMDC double layer.


\appendix

\section{Interactions in a two-component weakly interacting gas of A and B dipolar excitons}

\label{app:A}

In this appendix, we analyze the interactions in a two-component
weakly interacting gas of A and B dipolar excitons and derive the
system of equations for $R_{0i}$.

The chemical potentials $\mu_{A}$ and $\mu_{B}$ for A and B
excitons, respectively, of the weakly interacting Bose gas of
dipolar excitons within the Bogoliubov approximation are represented
as~\cite{BK}
\begin{eqnarray}
&& \mu_{A} -\mathcal{A}_{A}=g_{AA}n_{A}+g_{AB}n_{B} , \nonumber \\
 && \mu_{B} -\mathcal{A}_{B}=g_{BB}n_{B}+g_{AB}n_{A} ,
 \label{mu1}
\end{eqnarray}
where $n_{A(B)}$ is the concentration for A(B) dipolar excitons,
$g_{AA(BB)}$ and $g_{AB}$ are the exciton-exciton coupling
 constants for the interaction between two  A dipolar
excitons, two B dipolar excitons  and for the interaction between A
and B dipolar excitons, respectively, $\mathcal{A}_{A}$ and
$\mathcal{A}_{B}$ are the constants, determined by  A and B dipolar
exciton binding energy, correspondingly, and the gap, formed by a
spin-orbit coupling for the A (B) dipolar exciton~\cite{BK}.   One
can obtain from Eq.~(\ref{mu1}) the following equation:
\begin{eqnarray}  \label{mu3}
\mu_{A} + \mu_{B}
-\mathcal{A}_{A}-\mathcal{A}_{B}=g_{AA}n_{A}+g_{BB}n_{B}+g_{AB}n ,
\end{eqnarray}
where $n=n_{A}+n_{B}$ is the total 2D concentration of excitons.

Below we present the expressions for the coupling constant $g$ for
the both Keldysh and Coulomb potentials for the interaction between
the charge carriers.

According to the procedure presented in
Refs.~[\onlinecite{BK,BKKL}], two dipolar excitons cannot be closer
to each other than at the distance $R_{0i}$, which is determined by
the condition, following from the fact that the energy of two
dipolar excitons cannot exceed the doubled chemical potential $\mu $
of the system, i.e.~\cite{BK},
\begin{eqnarray}  \label{cond}
2\mathcal{A}_{A}+U(R_{0AA})=2\mu_{A}  ,\hspace{1.5cm}2\mathcal{A}%
_{B}+U(R_{0BB})=2\mu_{B} \ ,\hspace{1.5cm}\mathcal{A}_{A}+\mathcal{A}%
_{B}+U(R_{0AB})=\mu_{A} + \mu_{B}  ,
\end{eqnarray}
where $U(R)$ is the potential of interaction between the dipolar
excitons separated at the distance $R$. Using Eqs.~(\ref{mu1}),~(\ref{mu3}), and~(\ref{cond}), one obtains
\begin{eqnarray}   \label{geq1}
 \frac{U(R_{0AA})}{2} &=& g_{AA}n_{A}+g_{AB}n_{B} , \nonumber \\
 \frac{U(R_{0BB})}{2} &=& g_{BB}n_{B}+g_{AB}n_{A} , \nonumber \\
 U(R_{0AB}) &=& g_{AA}n_{A}+g_{BB}n_{B}+g_{AB}n  .
\end{eqnarray}%
Let us mention that the third  equation in Eq.~(\ref{geq1}) can be
replaced by the following:
\begin{eqnarray}   \label{geq13}
U(R_{0AB}) = \frac{1}{2} \left(U(R_{0AA}) + U(R_{0BB})\right) .
\end{eqnarray}%
Then the system of the equations~(\ref{geq1}) can be formed by the
first and the second equations from  Eq.~(\ref{geq1}) and
Eq.~(\ref{geq13}).

Substituting Eqs.~(\ref{Dipolar Keldysh Approx}) and~(\ref{gK}) into
Eq.~(\ref{geq1}), one obtains the following system of three
equations for $R_{0i}$ for the Keldysh potential:
\begin{eqnarray}   \label{geq1res}
&& 4\pi \rho _{0}^{2}y_{AA} \left[ n_{A}\left[
H_{0}\left(y_{AA}\right) -Y_{0}\left( y_{AA}\right) \right] +
n_{B}\left[ H_{0}\left(y_{AB}\right) -Y_{0}\left(y_{AB}\right)
\right] \right] = -\left[H_{-1}\left( y_{AA}\right) -
Y_{-1}\left(y_{AA}\right)
\right]  , \nonumber \\
&&  4\pi \rho _{0}^{2}y_{BB} \left[ n_{B}\left[
H_{0}\left(y_{BB}\right) -Y_{0}\left( y_{BB}\right) \right] +
n_{A}\left[ H_{0}\left(y_{AB}\right) -Y_{0}\left(y_{AB}\right)
\right] \right] = -\left[H_{-1}\left( y_{BB}\right) -
Y_{-1}\left(y_{BB}\right)
\right]  , \nonumber \\
&&  2 \pi \rho _{0}^{2}y_{AB} \left[ n_{A}\left[
H_{0}\left(y_{AA}\right) -Y_{0}\left( y_{AA}\right) \right] +
n_{B}\left[ H_{0}\left(y_{BB}\right) -Y_{0}\left( y_{BB}\right)
\right] + n \left[ H_{0}\left(y_{AB}\right)
-Y_{0}\left(y_{AB}\right)
\right] \right] \nonumber \\
&=& -\left[H_{-1}\left( y_{AB}\right) - Y_{-1}\left(y_{AB}\right)
\right]  ,
\end{eqnarray}
where  $y_{i} = R_{0i}/\rho_{0}$.

If the densities of A and B excitons are the same and
$n_{A}=n_{B}=n/2$, Eqs.~(\ref{geq1res})  result in the following
system of equations
\begin{eqnarray}   \label{geq1reseq}
&& 2 \pi n \rho_{0}^{2} y_{AA} \left[
H_{0}\left(y_{AA}\right)-Y_{0}\left(y_{AA}\right) +
H_{0}\left(y_{AB}\right)-Y_{0}\left(y_{AB}\right)  \right] = -
\left[H_{-1}\left(y_{AA}\right)-Y_{-1}\left(y_{AA}\right) \right] , \nonumber \\
&& 2 \pi n \rho_{0}^{2} y_{BB} \left[
H_{0}\left(y_{BB}\right)-Y_{0}\left(y_{BB}\right) +
H_{0}\left(y_{AB}\right)-Y_{0}\left(y_{AB}\right)  \right] = -
\left[H_{-1}\left(y_{BB}\right)-Y_{-1}\left(y_{BB}\right) \right] , \nonumber \\
&&  \pi n \rho_{0}^{2}  y_{AB} \left[
H_{0}\left(y_{AA}\right)-Y_{0}\left(y_{AA}\right) +
H_{0}\left(y_{BB}\right)-Y_{0}\left(y_{BB}\right) + 2
H_{0}\left(y_{AB}\right)- 2 Y_{0}\left(y_{AB}\right) \right]
\nonumber \\
&=&  - \left[H_{-1}\left(y_{AB}\right)-Y_{-1}\left(y_{AB}\right)
\right] .
\end{eqnarray}%

Substituting Eqs.~(\ref{Dipolar Keldysh Approx})
into~Eq.~(\ref{geq13}), one can replace the third equation in
Eqs.~(\ref{geq1res}) by the following equation:
\begin{eqnarray}   \label{geq1res3}
 \frac{1}{y_{AB}}\left[
Y_{-1}\left(y_{AB}\right) -H_{-1}\left( y_{AB}\right) %
\right] =  \frac{1}{2 y_{AA}}\left[
Y_{-1}\left(y_{AA}\right) -H_{-1}(y)\left( y_{AA}\right) %
\right]  + \frac{1}{2 y_{BB}}\left[
Y_{-1}\left(y_{BB}\right) -H_{-1}(y)\left( y_{BB}\right) %
\right] .
\end{eqnarray}%


\end{document}